\documentclass{aastex62}

\graphicspath{{./}{figures/}}

\accepted{July 1, 2020}
\submitjournal{AJ}

\begin{document}

\title{Transient Jupiter Co-orbitals from Solar System Sources}


\correspondingauthor{Sarah Greenstreet}
\email{sarah@b612foundation.org}

\author{Sarah Greenstreet}
\affil{Asteroid Institute,
20 Sunnyside Ave, Suite 427,
Mill Valley, CA 94941}
\affil{DIRAC Center,
Department of Astronomy,
University of Washington,
3910 15th Ave NE,
Seattle, WA 98195}

\author{Brett Gladman}
\affiliation{Department of Physics and Astronomy,
University of British Columbia,
6224 Agricultural Rd,
Vancouver, BC V6T 1Z1, Canada}

\author{Henry Ngo}
\affiliation{NRC Herzberg Astronomy \& Astrophysics Research Centre, 
5071 West Saanich Road, 
Victoria, BC V9E 2E7, Canada}

\begin{abstract}

We demonstrate dynamical pathways from main-belt asteroid and Centaur orbits to those in co-orbital motion with Jupiter, including the retrograde (inclination $i>90^o$) state. We estimate that at any given time, there should be $\sim1$ kilometer-scale or larger escaped asteroid in a transient direct (prograde) orbit with semimajor axis near that of Jupiter's ($a\simeq a_J$), with proportionally more smaller objects as determined by their size distribution. Most of these objects would be in the horseshoe dynamical state, which are hard to detect due to their moderate eccentricities (spending most of their time beyond 5 AU) and longitudes relative to Jupiter being spread nearly all over the sky. We also show that $\approx$1\% of the transient asteroid co-orbital population is on {\it retrograde} orbits with Jupiter. This population, like the recently identified asteroid (514107) 2015 BZ$_{509}$, can spend millions of years with $a\simeq a_J$ including tens or hundreds of thousands of years formally in the retrograde 1:-1 co-orbital resonance. Escaping near-Earth asteroids (NEAs) are thus likely the precursors to the handful of known high-inclination objects with $a\simeq a_J$. 
We compare the production of jovian co-orbitals from escaping NEAs with those from incoming Centaurs. We find that temporary direct co-orbitals are likely dominated by Centaur capture, but we only find production of (temporary) retrograde jovian co-orbitals (including very long-lived ones) from the NEA source. We postulate that the primordial elimination of the inner Solar System's planetesimal population could provide a supply route for a metastable outer Solar System reservoir for the high-inclination Centaurs.

\end{abstract}


\section{Introduction} \label{sec:intro}

A population of objects in co-orbital motion, as one of long-term stable and thus presumably primordial (i.e., $>4$~Gyr lifetimes) populations or as temporary captures, is known to exist with every planet in the Solar System with the sole exception of Mercury. Long-range planetary interaction can cause an object with semimajor axis very close to the planet to oscillate around the L4 or L5 Lagrange point (called trojan motion), around a point 180$^\circ$ away from the planet (called horseshoe motion), or even around the planet's longitude (quasi-satellites). Earth currently has a population of five horseshoe \citep{Wiegertetal1998,ChristouAsher2011,dlFM22016b}, five quasi-satellite \citep{Connorsetal2004,Wajer2010,dlFM22016c}, one Trojan \citep{Connorsetal2011}, and four horseshoe/quasi-satellite combination \citep{Connorsetal2002,Brasseretal2004,dlFM22016a} co-orbitals, all of which are on orbits unstable on timescales much shorter than the Solar System's age. Venus has been discovered to temporarily host: one quasi-satellite \citep{Mikkolaetal2004}, one Trojan \citep{dlFM22014a}, one quasi-satellite/horseshoe complex co-orbital \citep{Brasseretal2004}, and one Trojan/horseshoe combination co-orbital \citep{dlFM22013b}. A total of eight long-term stable Trojans have been discovered to co-orbit Mars \citep{Scholletal2005,dlFM22013a}.

Among the giant planets, Jupiter and Neptune are known to have large stable Trojan populations, the Neptune Trojans possibly outnumbering those of Jupiter \citep{Alexandersenetal2016}. Neptune has also been discovered to have a handful (8 in total so far) of temporarily-trapped Trojans on unstable orbits \citep{Brasseretal2004,HornerLykawka2010,dlFM22012a,dlFM22012b,Guanetal2012,HornerLykawka2012,Horneretal2012,Alexandersenetal2016}. Uranus has two known temporary Trojans \citep{Alexandersenetal2013,dlFM22017}. Saturn was very recently discovered to host four possible transient co-orbitals on retrograde (inclination $i>90^o$) orbits with very short (i.e., $\lesssim4$~kyr) potential captures \citep{MoraisNamouni2013a,Lietal2018}. In addition, Jupiter may have a few very short-term ($<1$~kyr) captured co-orbitals \citep{Karlsson2004}.

The longest-lived transient co-orbital discovered to date for Jupiter and Saturn is a co-orbital with Jupiter \citep{Wiegertetal2017,NamouniMorais2018}; (514107) 2015 BZ$_{509}$ (hereafter referred to as BZ509) is currently on a retrograde jovian co-orbital orbit ($a=5.139$~AU, $e=0.380$, $i=163.022^o$, $H=16.0$) and remains resonant for tens of thousands of years \citep{Wiegertetal2017}. BZ509 was shown by \citet{Wiegertetal2017} in addition to remain with semimajor axis $a$ near (within a few tenths of an AU) that of Jupiter for $\sim1$~Myr, with often no formal resonant angle libration. \citet{Huangetal2018} show additional integrations with libration of the resonant argument for $\approx200$~kyr. To study the long-term stability of BZ509, \citet{NamouniMorais2018} numerically integrated one million clones of the object and found a 0.003\% chance that $a$ remains near that of Jupiter for $>4$~Gyr. Citing the Copernican Principle that posits BZ509 has not been observed at any preferred epoch in Solar System history, \citet{NamouniMorais2018} proposed BZ509 is an interstellar object that was captured into the retrograde jovian co-orbital state $>4$~Gyr ago. However, we demonstrate it is also possible a population of temporarily-stable jovian retrograde co-orbitals are continuously resupplied from a source within the Solar System. Such a steady-state resupply source has been used \citep{Alexandersenetal2013} to successfully explain the number of transient neptunian Trojans and the single temporarily-trapped uranian Trojan 2011 QF99.

\section{Potential sources of jovian co-orbitals} \label{sec:sources}

We consider two potential Solar System sources for transient jovian co-orbitals on direct and retrograde orbits: the near-Earth asteroids and inwardly-migrating Centaurs. Using the two models described below, we search for transient jovian co-orbital production (on both direct and retrograde orbits) from each source population and estimate their steady-state population.

The near-Earth object (NEO) orbital distribution model from \citet{Greenstreetetal2012a} provides the steady-state NEO population originating from escaping main-belt asteroids. The orbital histories of 7,000 test particles integrated for 100-200 Myr were stored at 300 year intervals. The vast majority of the asteroid test particles were captured into the NEO region, but some migrated outward in semimajor axis from the main asteroid belt, often getting ejected from the Solar System by Jupiter. \citet{Greenstreetetal2012a,Greenstreetetal2012b} discovered that 0.2\% of the steady-state NEO population is on retrograde (inclination $i>90^o$) orbits. We search the orbital histories of all NEO model test particles for temporarily-trapped jovian co-orbitals in both direct and retrograde orbits.

A steady-state model of the $a<34$~AU Centaur population as computed by \citet{Alexandersenetal2013}, using an incoming scattering object model from \citet{Kaibetal2011}, was used to determine the frequency of temporarily-trapped co-orbitals on direct orbits with Uranus and Neptune. This model was updated by \citet{Alexandersenetal2018} to extend to transient co-orbitals of Saturn and a lower-limit on those with Jupiter. The orbital histories for all test particles were stored at 50 year intervals for a total integration time of 1~Gyr. In addition to the the temporarily-trapped co-orbitals with Jupiter on direct orbits searched for in \citet{Alexandersenetal2018}, we search the Centaur histories for transient jovian co-orbitals on retrograde orbits.

We have not included the Oort cloud as a potential source region. Although it is possible for Oort cloud comets with small enough perihelia to have their aphelia dropped to within the giant planet region through numerous planetary close encounters, the efficiency of this process is likely low. In any case, such objects would almost certainly first transit through the moderate-$a$ state that is our source region, and thus if some TNOs in that region are returning Oort cloud objects they are already included in our model.

\subsection{Co-orbital Detection} \label{subsec:coorb_detection}

The formal definition of a direct co-orbital state is that the resonant angle $\phi_{1:1}=\lambda-\lambda_{planet}$ librates, where $\lambda$ is the mean longitude of the small body and $\lambda_{planet}$ is the mean longitude of the planet. While detecting this libration in the 0.5~TB of NEO orbital histories (at 300 year intervals) and the 250~GB of Centaur orbital histories (at 50 year intervals) is difficult to automate, an automatic process is necessary to filter the large outputs.

Instead, to diagnose whether particles are co-orbital we used the simpler method of scanning the semimajor axis history using a running window, which diagnoses co-orbitals well \citep{Alexandersenetal2013}. The length of the running window was chosen in each source region's case (NEOs: 9~kyr, Centaurs: 5~kyr) to be several times longer than the typical libration period at Jupiter. A particle was classified as a co-orbital if, within the running window, both its average semimajor axis $a$ was within 0.4~AU of Jupiter's average $a$ and no individual semimajor axis value differed by more than 3.5 times Jupiter's Hill-sphere radius $R_H=1.2$~AU from Jupiter's $a$. If these requirements were met, the orbital elements and the integration time at the center of the running window for that particle were output to indicate co-orbital motion in that window. The window center was then advanced by a single integration output interval (300 years for the NEOs and 50 years for the Centaurs) and the diagnosis was performed again on the next running window. This records consecutive identifications of a particle temporarily trapped in co-orbital motion with Jupiter as a single ``trap" until the object is scattered away. A minor shortcoming of this co-orbital identification method is that the beginning and end of each trap is not well-diagnosed due to the ends of the window not entirely falling withing the trap at these times. This method provides us with estimates of the duration of temporary traps, each of which must be greater than the length of the running window to be diagnosed, to within a factor of two accuracy \citep{Alexandersenetal2013}.

\subsection{Resonant Island Classification} \label{subsec:res_island_class}

For each time step a particle has been classified as a co-orbital, we determine in which of the four resonant islands the particle is librating, i.e., whether it is a horseshoe, L4 Trojan, L5 Trojan, or quasi-satellite, using a method similar to that in \citet{Alexandersenetal2013}. Our co-orbital detection algorithm produced nearly 1,800 total temporary traps, which requires another automated process to determine resonant island classification. As with the detection algorithm, this is similarly difficult to automate especially because complex variations and combinations can exist for high inclinations.

For our resonant island classification algorithm, we examine the behavior of two versions of the resonant angle $\phi_{1:1}$. For objects in direct co-orbital motion with Jupiter, we use the traditional definition of the resonant angle $\phi_{1:1}=\lambda-\lambda_J$. If $\phi_{1:1}$ remains in the leading or trailing hemisphere for the duration of a running window, we assign the particle to the L4 or L5 state, respectively. If $\phi_{1:1}$ crosses $180^o$ at any time during the window interval, the co-orbital is labelled a horseshoe. All remaining orbits are classified as quasi-satellites, as they must be co-orbitals that cross between the leading and trailing hemispheres at $\phi_{1:1}=0^o$ and not at $180^o$. 

Although the possibility of erroneous classifications exist with this method, we find these errors affect $<10\%$ of cases upon manual inspection of dozens of cases. The majority of these examples were particularly chosen as co-orbitals that experience multiple transitions between Trojan, horseshoe, and/or quasi-satellite states as well as possible times of non-resonant behavior (resonant argument circulation) as the temporary co-orbitals move in and out of 1:1 resonant capture. To ensure accurate classification of periods of resonant argument libration within a running window, the average and individual semimajor axis limits and running window length described in Section~\ref{subsec:coorb_detection} above were adjusted until periods of co-orbital behavior with resonant argument libration were correctly identified $>90\%$ of the time. In addition, these parameters were adjusted to increase correct classifications of resonant island libration behavior (i.e., Trojan, horseshoe, and quasi-satellite behavior) to the same level of accuracy. This includes periods of transitions between multiple resonant islands, which almost always occur on timescales longer than the length of the running window. An additional minor shortcoming of this co-orbital identification method is the difficulty of correctly classifying resonant island libration during periods of transition between states, however, as stated above, we find these affect $<10\%$ of co-orbital classifications in our simulations. We note that another limitation to our resonant island classification method is that co-orbitals with large amplitude librations that encompass libration around Lagrange points not typically associated with their resonant state (e.g., large amplitude Trojans whose librations extend beyond either the leading or trailing hemisphere to $\phi>180^o$ or $\phi<0^o$) would likely not be well classified with our identification method. However, we find such large amplitude libraters to be rare ($<10\%$) among the transient co-orbitals in our simulations. Thus, inaccurate classifications do not greatly affect our co-orbital fraction and resonant island distribution estimates, supporting our goal of better than factor of two accuracy.

Following the convention for retrograde orbits of \citet{MoraisNamouni2013b}, we define the 1:-1 resonant argument for retrograde orbits to be $\phi^\star=\lambda^\star-\lambda_J-2\omega^\star$ (their Equation 9). Here, $\lambda_J$ is the mean longitude of Jupiter and is defined in the usual planetary sense of being measured always along the direction of orbital motion. $\lambda^\star$ is the mean longitude of the particle and is defined as $\lambda^\star=M+\omega-\Omega$, where $\Omega$ is the longitude of ascending node measured in the planetary sense from the reference direction, $\omega$ is the argument of perihelion measured from the ascending node to the pericenter in the direction of motion (opposite the direction of the measured angle $\Omega$ for retrograde orbits), and $M$ is the mean anomaly also measured along the direction of motion. Lastly, $\omega^\star$ is the particle's longitude of perihelion and is defined as $\omega^\star=\omega-\Omega$, where $\omega$ and $\Omega$ are defined above. This expression for the 1:-1 resonant argument reduces to the equation found in \citet{NamouniMorais2018}, which is written as $\phi^\star=\lambda-\lambda_{J}-2\omega$, where $\lambda$ is the particle's mean longitude and defined as $\lambda=M+\omega+\Omega$.

Similar to the method described above for direct jovian co-orbitals, for retrograde jovian co-orbitals we examine the behavior of both the traditional resonant angle ($\phi_{1:-1}=\lambda-\lambda_J$) and $\phi^\star_{1:-1}=\lambda^\star-\lambda_J-2\omega^\star$ above \citep{MoraisNamouni2013b}. It is important to note that in the retrograde co-orbital case, the traditional interpretation of the resonant island around which a co-orbital librates is not relevant. For example, in the case that $\phi_{1:-1}$ librates around $0^o$, the co-orbital does not appear to orbit the planet in the co-rotating frame as in the ``quasi-satellite" direct case. Rather, the co-orbital and the planet move in {\it opposite} directions with the same mean-motion keeping $\phi_{1:-1}$ near zero, but their opposing trajectories result in them not remaining near each other.

\section{Example Temporary Co-orbital Traps}
\label{sec:example traps}

Temporary jovian co-orbitals can be captured from either asteroids migrating outward toward Jupiter or Centaurs migrating inward toward Jupiter. In this section we discuss the typical dynamical behavior of these transient co-orbitals, including their orbital evolutions and typical eccentricity and inclinations from both the asteroidal and Centaur sources. 

\begin{figure}[h!]
\centering
\includegraphics[width=0.65\textwidth]{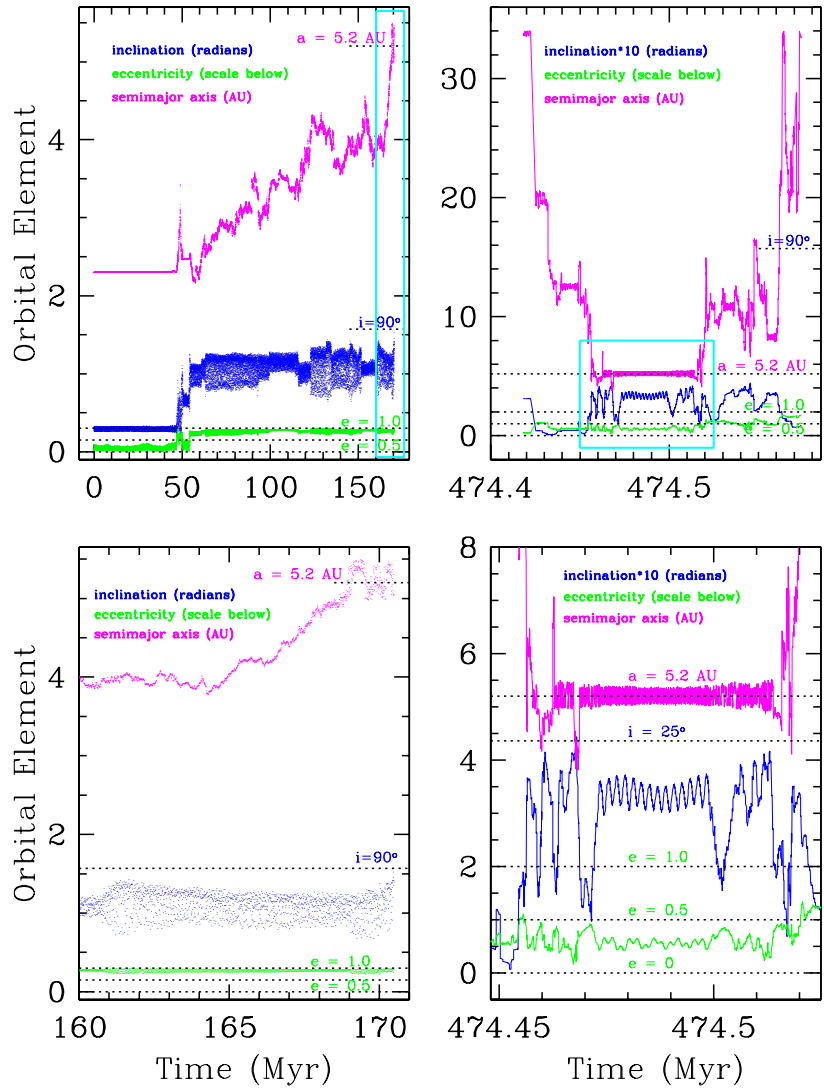}
\caption{Two examples of direct jovian co-orbital temporary captures. Top Left: Example orbital evolution of a temporary jovian NEO co-orbital on a direct orbit. The trap lasts for $\approx3$~Myr. The cyan box marks the region of the zoom-in (bottom left) around the time of co-orbital capture. Bottom Left: Zoom-in of the time around the $\approx3$~Myr temporary jovian NEO co-orbital capture. Top Right: Example orbital evolution of a temporary jovian Centaur co-orbital on a direct orbit. This is the longest-lived transient jovian Centaur co-orbital found in the simulations, captured for a consecutive 45~kyr with a brief 5~kyr capture a few thousand years earlier. The cyan box marks the zoomed-in region (bottom right) around the time of co-orbital capture. Bottom Right: Zoom-in of the time around the temporary jovian Centaur co-orbital capture lasting for 45~kyr preceded by a brief 5~kyr capture a few kyr earlier. Note the inclination scales are in radians with a reference level (in degrees) indicated.  \label{fig:a_e_i_NEO_Centaur_direct}}
\end{figure}

Figure~\ref{fig:a_e_i_NEO_Centaur_direct} shows example orbital evolutions for direct jovian co-orbital captures from an asteroidal source (left top \& bottom panels) and a Centaur source (right top \& bottom panels). The captured asteroid co-orbital leaves the $\nu_6$ resonance source $\approx50$~Myr into its lifetime. It then random walks in $a$ for the next $\approx120$~Myr, during which time it experiences Kozai oscillations in $e$ and $i$ at high-$e$ and high-$i$ (though still on a direct orbit with $i<90^o$). At $\approx170$~Myr into the particle's lifetime, it becomes temporarily captured as a direct jovian co-orbital. The trap lasts for $\approx3$~Myr before the perihelion drops to the solar radius.

The example Centaur jovian co-orbital capture (right panels of Figure~\ref{fig:a_e_i_NEO_Centaur_direct}) only enters the $a<34$~AU region after the first 474.4~Myr of its lifetime. It then quickly drops from transneptunian space to $a\simeq$~$a_J$ in $\approx30$~kyr and remains with $a$ near that of Jupiter for $\approx55$~kyr. The co-orbital trap lasts for $\approx45$~kyr (this is the longest of all the Centaur jovian co-orbital captures found) with a brief 5~kyr trap a few thousand years earlier. The inclination never reaches more than $\simeq20^o$ throughout the particle's time with $a<34$~AU. The semimajor axis then random walks back out to transneptunian space over the next $\approx165$~kyr.

\begin{figure}[h!]
\centering
\includegraphics[width=0.35\textwidth,angle=270]{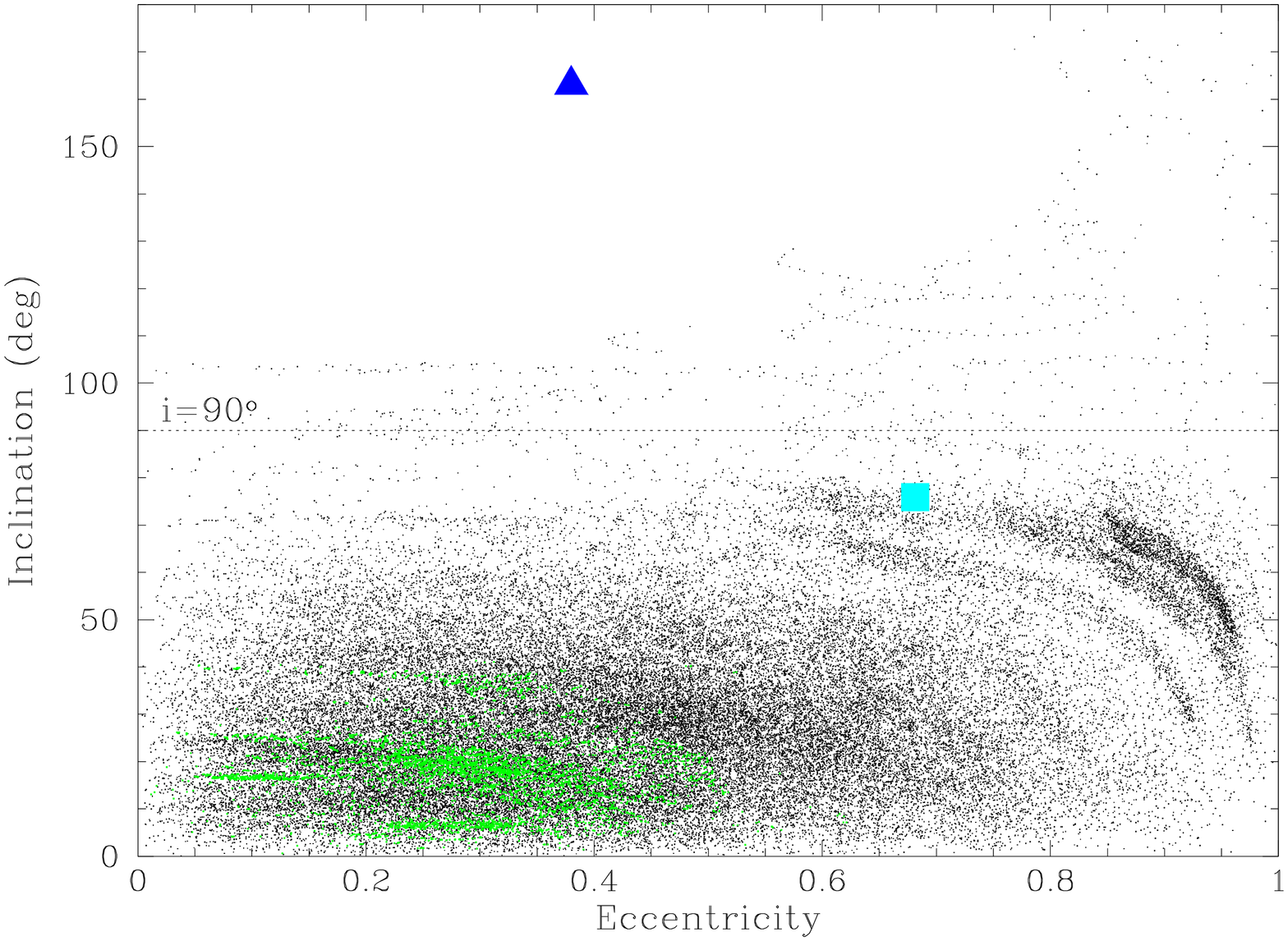}
\caption{Inclination (deg) vs eccentricity for each time step that all asteroidal particles have semimajor axes near that of Jupiter thinned by a factor of three for visibility (black dots). The cyan square marks the $e$ and $i$ of the cloned asteroid particle at the instance of cloning (see Section~\ref{sec:retro} below). The blue triangle indicates BZ509's current $e$ and $i$. The green points show $e$ and $i$ of the $a<34$~AU, $q>2$~AU Centaurs that become temporarily-trapped jovian co-orbitals. \label{fig:e_vs_i_asteroids_Centaurs}}
\end{figure}

The eccentricity and inclination behavior of asteroids migrating outward from the main asteroid belt to semimajor axes near that of Jupiter and Centaurs migrating inward toward Jupiter is shown in Figure~\ref{fig:e_vs_i_asteroids_Centaurs}. Asteroids with $a$ near that of Jupiter explore all values of eccentricity ($e$) and inclinations $i<90^o$ (direct orbits). In addition, a handful of particles in the NEO model \citep{Greenstreetetal2012a} reach $i>90^o$ while $a\simeq$~$a_J$; those particles with $i>90^o$ visit the full range of possible eccentricities from $0-1$ (Section~\ref{sec:retro} discusses this in greater detail).

Centaurs with $a<34$~AU and $q>2$~AU evolving inward to semimajor axes near $a_J$ are found to be confined to $e<0.6$ and $i<50^o$. Figure~\ref{fig:e_vs_i_asteroids_Centaurs}  shows a subset of Centaurs that include all the temporary jovian co-orbital captures as well as shorter total durations explored with $a$ near that of Jupiter in the simulations. The $e\lesssim0.6$ cut for $a=5.2$~AU is due to the $q>2$~AU cut in the simulations. We find no $i>50^o$ Centaurs with $a\approx$~$a_J$ (see Section~\ref{sec:discussion} for more discussion).

\section{Temporary Co-orbital Time Scales}
\label{sec:timescales}

\begin{figure}[h!]
\centering
\includegraphics[width=0.5\textwidth,angle=270]{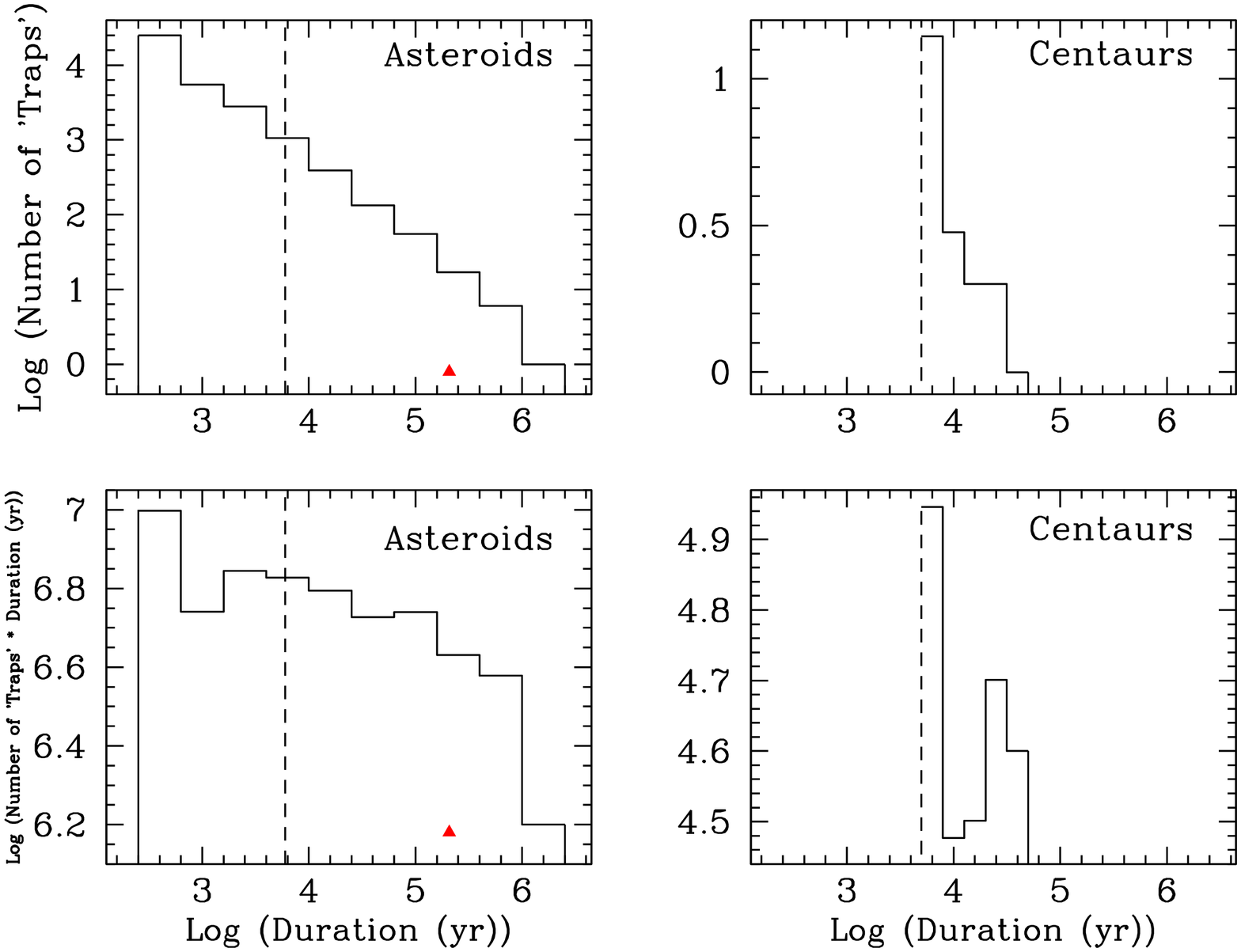}
\caption{
Duration of near-Jupiter residence, presented as histograms of times that particles have $a\simeq$~$a_{J}$. TOP: Logarithm of the number of co-orbital ``traps" of each duration observed from an asteroidal (left) and Centaur (right) source. The dashed lines mark the amount of time particles must remain with $a$ near $a_{J}$ for each source region (9~kyr for asteroids, 5~kyr for Centaurs) to be classified as co-orbitals. The top left panel shows all ``traps" (i.e., consecutive time steps) with $a\simeq$~$a_{J}$ for the asteroidal source, where the traps to the right of the vertical dashed line are those that we classify as temporary jovian co-orbital traps. The top right panel only shows the traps for the classified temporary jovian co-orbitals from the Centaur source. Long total durations with $a\simeq$~$a_J$ represent single particles that get trapped for long contiguous time periods, but shorter-duration traps are more numerous. The red triangle corresponds to the asteroidal co-orbital trap event for one of the retrograde captures (shown in Figure~\ref{fig:a_e_i_NEO_retro_and_zoom}). BOTTOM: Time weighted residence, showing likelihood of finding a particle resident for that duration. For example, asteroidal particles (left) with 30 kyr traps occur three times less than 10 kyr traps (upper plot) but when time weighted by the trap durations should be roughly equally likely to be found (bottom plot). The longest-lived temporary jovian NEO co-orbital capture lasts for 2.4~Myr. Right: Centaur source. The result is that most observed temporary jovian Centaur co-orbitals should have short trap durations of 5-8~kyr with longer resonant captures being a factor of 2-3 less likely to be found. The longest-lived temporary jovian Centaur co-orbital capture lasts for $\approx45$~kyr.
\label{fig:duration_e_i_NEOs_Centaurs}
}
\end{figure}

Asteroidal particles visit semimajor axes near that of Jupiter for durations ranging from 300 years (our minimum sampling, which is visible in the smallest bin in the left two panels of Figure~\ref{fig:duration_e_i_NEOs_Centaurs}) to a maximum 2.4~Myr, although the majority of times in this area of phase space fall between 2~kyr and 100~kyr. The shortest durations in this range are due to a few integration sampling intervals when particles quickly pass through the phase space near Jupiter on their way from the main belt to the outer Solar System. As described in Section~\ref{subsec:coorb_detection} above, however, asteroidal particles must remain with $a$ near $a_{J}$ for 9,000 years to be classified as co-orbitals. We find the mean, median, and maximum lifetimes for transient asteroidal jovian co-orbitals on direct orbits are 25~kyr, 14~kyr, and 2.4~Myr, respectively.

As shown in Figure~\ref{fig:duration_e_i_NEOs_Centaurs}, long total durations with $a\simeq$~$a_J$ represent single particles that get trapped for long contiguous time periods, but shorter-duration time periods are more numerous. Asteroidal particles with 30~kyr co-orbital traps occur three times less than 10~kyr traps (upper left panel of Figure~\ref{fig:duration_e_i_NEOs_Centaurs}), but when time weighted by the trap durations should be roughly equally likely to be found (bottom left plot of Figure~\ref{fig:duration_e_i_NEOs_Centaurs}). 

Transient jovian Centaur co-orbitals must remain with $a\simeq$~$a_{J}$ for 5~kyr to be classified as co-orbitals. The majority of the temporarily-trapped jovian co-orbital captures from the Centaur source last between  5-8~kyr (right two panels of Figure~\ref{fig:duration_e_i_NEOs_Centaurs}). Thus, most observed temporary jovian Centaur co-orbitals should have short trap durations of 5-8~kyr with longer resonant captures being a factor of 2-3 less likely to be found. We find the mean, median, and maximum lifetimes for transient Centaur jovian co-orbitals on direct orbits are 11~kyr, 7.4~kyr, and 45~kyr.

\section{Temporary Co-orbital Population Estimates}
\label{sec:pop estimates}

We find that 0.11\% of the steady-state NEO population are temporarily-trapped jovian co-orbitals on direct orbits. Given that there are $\simeq1,000$ NEOs with $H<18$, this means we would expect there to be one transient jovian co-orbital on a direct orbit trapped from the population of main-belt asteroids at any time (with more smaller ones as expected from whatever the unknown size distribution is).

The larger population of Centaurs (compared to the NEA population) means Centaur capture into temporary direct jovian co-orbitals might outnumber those captured from the NEA population. Our simulations indicate that 0.001\% of the $a<34$~AU, $q>2$~AU Centaurs in steady-state are temporarily-trapped jovian co-orbitals on direct orbits, a roughly two orders of magnitude smaller fraction. It is estimated, however, that there are between $\sim1\times10^6$ and $\sim4\times10^6$ $H<18$, $a<34$~AU Centaurs \citep{Lawleretal2018}, and so this $\sim 1000\times$ larger population results in an expectation of $\sim10-40$ direct transient jovian co-orbitals from a Centaur source at any given time. (We point out that the $q>2$~AU boundary in the simulations may mean this number is actually larger than this estimation, although we find that only roughly 5\% of the particles are discarded from the simulations because they get within this inner distance cut.) The uncertainty on this estimate is at least an order of magnitude\footnote{Not only because of uncertainties in the size distribution, but also because as $a=5$~AU is approached small Centaurs may be modified by splitting \citep{Fernandezetal2009}.}, meaning the number ratio of transient jovian co-orbitals on {\it direct} orbits coming from the Centaur and asteroidal sources is $\sim$1 -- 100.

Our results show that there must be temporarily-trapped direct jovian co-orbitals with lifetimes of $10^4$--$10^6$ years, but none have ever been reported. Using $\pm$1,000-year integrations, \citet{Karlsson2004} studied a handful of $<1$~kyr temporary captures in the known candidate Jupiter Trojan population, but none of these objects are metastable for the much longer timescales we diagnose here. Identification of such transient jovian co-orbitals will eventually happen (just as such orbits have been identified for the other giant planets) but is challenging for a number of reasons. Firstly, given the population of the NEA and Centaur sources, we expect such objects to be not much brighter than $H\sim18$ and thus be faint; there are essentially no multi-opposition orbits in the Minor Planet Center database with $H>17$. Secondly, our simulations show that the majority of these temporary traps have $e>0.3$ and thus spend a much larger fraction of their time far from the Sun beyond survey detection limits. Thirdly, the majority of these objects are horseshoe or quasi-satellite orbits (Table~\ref{table:res_island_breakdown_direct}) and thus are not confined near the Lagrange points and thus their co-orbital semimajor axes might not even be recognized in a very short arc orbit discovered at heliocentric distances between 4 and 7~AU. Even once recognized, it will require a very precise orbit to confirm the co-orbital behavior as the orbital uncertainty needs to be shrunk so much that {\it all} orbits that fit the observations show co-orbital behavior \citep{Alexandersenetal2013}; this is a difficult standard to surpass, but our results indicate that once this is done there should be some small direct co-orbital objects that librate securely in the 1:1 resonance for tens to hundreds of librations before leaving.
		
\begin{table}[h!]
\begin{center}
\begin{tabular}{| c | c | c | c |}
\hline
\bf{Source} & \bf{Horseshoes} & \bf{Trojans} & \bf{Quasi-satellites}\\
\hline
Asteroids & 93\% & 2\% & 5\% \\
Centaurs & 59\% & 21\% & 20\% \\
\hline
\end{tabular}
\end{center}
\caption{Resonant island classifications for transient jovian co-orbitals on direct orbits from the asteroidal and Centaur sources. Note that the quasi-satellites either outnumber or roughly equal the Trojans, but our resonant island classification method, as described in Section~\ref{subsec:res_island_class}, may overestimate the fraction of quasi-satellites.}
\label{table:res_island_breakdown_direct}
\end{table}

\section{Retrograde Jovian Co-orbital Dynamics} \label{sec:retro}

In addition to finding transient jovian co-orbitals on direct orbits from among particles in the NEO model, we observe a handful of particles evolving to {\it retrograde} ($i>90^o$) co-orbitals. Some of these particles evolve to $a\simeq$~$a_J$ then flip to retrograde orbits (three particles), and some flip to retrograde orbits before going to Jupiter (five particles). For particles classified as co-orbitals (see Section~\ref{subsec:res_island_class} for definition) we find mean, median, and maximum times with consecutive running window centers with $i>90^o$ of 11~kyr, 7~kyr, and 87~kyr, respectively. These temporarily-trapped jovian co-orbitals on retrograde orbits represent 0.001\% of the steady-state NEO population. Given that there are $\simeq1,000$ $H<18$ NEOs, this means we would expect there to be one transient jovian co-orbital on either a direct or retrograde orbit trapped from the asteroid population (with more smaller ones) and that co-orbital would have a 1\% chance of existing on a retrograde orbit. 

\begin{figure}[h!]
\centering
\includegraphics[width=0.45\textwidth]{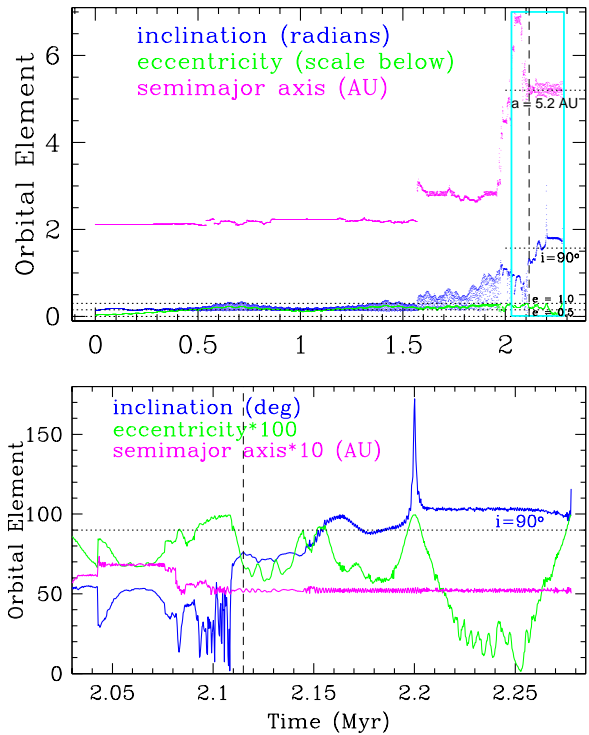}
\caption{Retrograde asteroidal jovian co-orbital example. Top: Orbital history of a particle from our NEO model integrations of an asteroid that becomes a jovian co-orbital around the time of its flip to a retrograde orbit. This particle was cloned 9,900 times around 2.115~Myr into its lifetime (indicated by vertical dashed line) to search for additional retrograde jovian NEO co-orbital behavior. The cyan box marks the zoomed-in region shown in the bottom panel. Bottom: Zoomed-in version of top panel beginning just before the particle becomes locked with its semimajor axis near that of Jupiter and the particle then flips to a retrograde orbit ($i>90^o$). The vertical dashed line marks the time at which the particle was cloned. \label{fig:a_e_i_NEO_retro_and_zoom}}
\end{figure}

Figure~\ref{fig:a_e_i_NEO_retro_and_zoom} shows the full orbital evolution of a particle from our NEO model that becomes trapped as a retrograde jovian co-orbital. This particle lives in the $\nu_6$ resonance for $\approx1.5$~Myr before experiencing a series of planetary close encounters that eventually kick its semimajor axis exterior to Jupiter before it drops back to $a\approx$~$a_J$ around 2.1~Myr into its lifetime. It then remains a co-orbital for $\approx180$~kyr. Shortly after reaching $a\approx$~$a_J$, the inclination becomes retrograde. The particle then remains a retrograde jovian co-orbital for a total of $\approx130$~kyr with the longest consecutive retrograde co-orbital period lasting for the final 100~kyr of the particle's lifetime. While the particle is on a retrograde orbit $e$ explores nearly all possible values. At $\approx2.2$~Myr into the particle's lifetime, a brief spike in $e$ and $i$ occurs with $e$ reaching\footnote{The 4-hour time step in the NEO model integrations satisfies the needed time step to resolve solar encounters (and detect collisions), of less than $P(1-e^2)^{3/2}/3$ = 36 hours, where $P$ is the orbital period, \citep{RauchHolman1999} by about an order for magnitude at $e=0.995$.} nearly one ($e\approx0.995$; $q\approx0.026$) and $i$ exceeding $170^o$. The inclination then drops and settles at $i\simeq100^o$ while $e$ plummets to nearly zero 50~kyr later before climbing to $e=1$ ($e>0.999$) at approximately 2.28~Myr into the particle's lifetime when it is pushed into the Sun ($r<0.005$~AU).

The only path we have been able to demonstrate to the retrograde state is from the main asteroid belt source. As reported in Section~\ref{sec:example traps}, we found no examples of incoming Centaurs reaching temporarily-trapped retrograde jovian co-orbitals with $q>2$~AU. This is likely a result of the different inclination distributions of asteroids and Centaurs that reach the $a\simeq$~$a_J$ region. Although some asteroids get captured into the jovian co-orbital state from retrograde orbits, those that are captured as co-orbitals on direct orbits that then flip to retrograde as well as those that remain on direct orbits are captured at significantly higher inclinations (up to $i=60^o-80^0$) than Centaurs upon capture ($i<35^o$). The higher direct-orbit inclinations of asteroidal particles in the $a\simeq$~$a_J$ region likely give the asteroids an advantage over the Centaurs for reaching the retrograde co-orbital state.

Figure~\ref{fig:e_vs_i_asteroids_Centaurs} showed $e$ vs $i$ for output intervals when $a$ is near that of Jupiter for the asteroidal source. The blue triangle in Figure~\ref{fig:e_vs_i_asteroids_Centaurs} marks the current $e$ and $i$ of BZ509. We do not see any particles in the NEO model reaching both this $e$ and $i$ at the same time, although these values are reached independently by particles in the model. Because we only find eight particles in the NEO model that become transient retrograde jovian co-orbitals, none of which have $e$ and $i$ simultaneously near that of BZ509, we cloned the particle shown in Figure~\ref{fig:a_e_i_NEO_retro_and_zoom} to attempt to find particles reaching these eccentricities and inclinations simultaneously; this also provided a suite of retrograde jovian co-orbital examples to better understand their typical behavior. The cyan square in Figure~\ref{fig:e_vs_i_asteroids_Centaurs} shows the $e$ and $i$ at cloning for the cloned particle. This particle was cloned at $\approx2.115$~Myr into its lifetime while the particle is classified as a jovian co-orbital, but shortly {\it before} it became retrograde (see Figure~\ref{fig:a_e_i_NEO_retro_and_zoom} for the detailed orbital evolution of the particle).

\begin{figure}[h!]
\centering
\includegraphics[width=0.35\textwidth,angle=270]{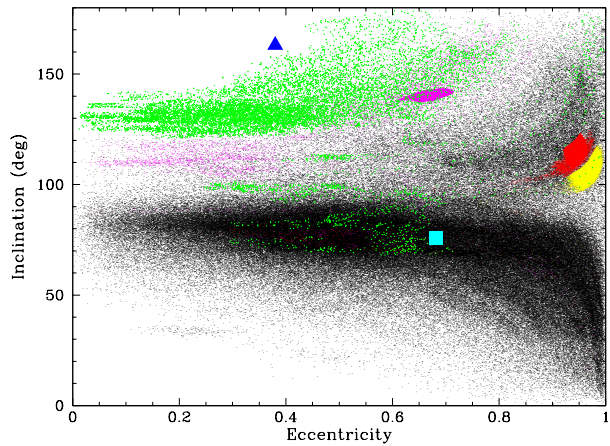}
\caption{Inclination vs eccentricity range for evolutions departing from the cloned particle (cyan square, and see Fig.~\ref{fig:a_e_i_NEO_retro_and_zoom}). For 5\% of the time steps that clone particles (reduced to prevent figure saturation) have semimajor axes near Jupiter, a black dot is plotted. The blue triangle indicates BZ509's current $e$ and $i$. (The blue triangle and cyan square are the same as in Figure~\ref{fig:e_vs_i_asteroids_Centaurs}). The green dots show the $e$/$i$ evolution for the full orbital history of the clone particle shown in Figure~\ref{fig:clone_example}, which is the closest particle to simultaneously matching the current $a$, $e$, \& $i$ of BZ509. The magenta, red, and yellow points show the $e$/$i$ evolution for the full orbital history of single particles with near-Jupiter visits as long as BZ509 (see Appendix). See text for discussion.
\label{fig:all_e_vs_i_clones}
}
\end{figure}

\begin{figure}[h!]
\centering
\includegraphics[width=0.45\textwidth]{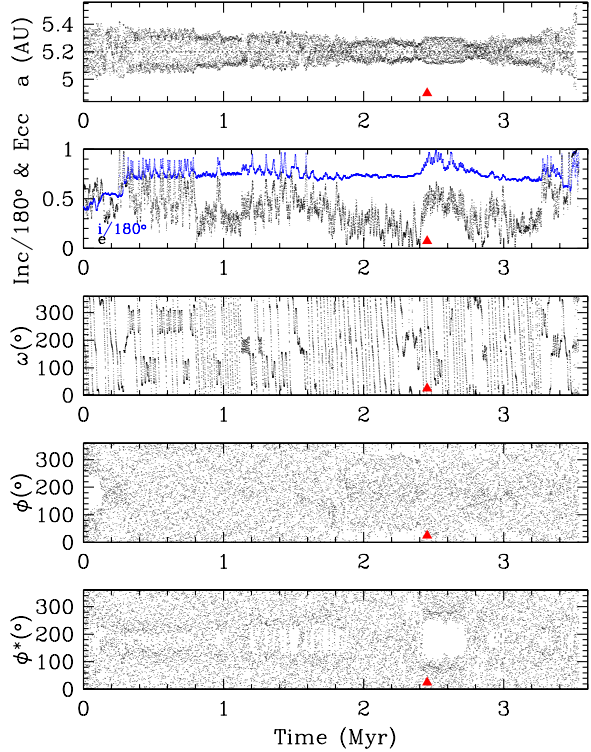}
\caption{Orbital history of a clone of the particle shown in Figure~\ref{fig:a_e_i_NEO_retro_and_zoom} starting just before it flips to a retrograde state at $\approx0.2$~Myr. The particle spends $\approx3.52$~Myr (shown here) with $a$ near Jupiter's semimajor axis before its semimajor axis suddenly drops to near $a\approx3$~AU after a planetary close encounter and it collides with Jupiter $\approx16$~kyr later. The particle flips to a retrograde orbit $\approx30$~kyr into its lifetime and coupled Kozai $e$ and $i$ oscillations often occur while the particle is on a retrograde orbit as well as the argument of pericenter ($\omega$) sometimes librating around either $90^o$ or $270^o$ during this time. The resonant arguments $\phi_{1:-1}=\lambda-\lambda_J$ and $\phi^\star_{1:-1}=\lambda-\lambda_J-2\omega$ \citep{NamouniMorais2018} show resonant behavior (libration around $180^o$ and $0^o$, respectively) both before and after the flip to a retrograde orbit, indicating the particle is at times in the 1:-1 co-orbital resonance with Jupiter. This clone particle also spends a single 300 year dump interval within 0.35 AU, $e=0.05$, and $i=3^o$ of 2015 BZ$_{509}$'s current elements. The red triangle at $\approx2.45$~Myr indicates the time when the cloned particle's orbit is closest to that of BZ509 ($a\approx5.4$~AU, $e\approx0.38$, $i\approx166^o$). \label{fig:clone_example}}
\end{figure}

We cloned this particle 9,900 times by randomly `fuzzing' the position and velocity vector components to be within $\pm5$x$10^{-9}$~AU (0.75~km) and $\pm5$x$10^{-9}$~AU/yr (0.75~km/yr) of their initial values, respectively. We then performed integrations for 10~Myr, by which time 99.5\% of the particles had been removed. Figure~\ref{fig:all_e_vs_i_clones} shows the $e$ vs $i$ plot for 5\% of these clones (black dots; thinned for better visibility) as well as the $e$/$i$ values for the full orbital history of the clone particle (green dots) that comes closest to matching BZ509's $a$, $e$, and $i$ as well as three additional long-lived clone particles with $a$ near $a_J$ (magenta, red, and yellow dots). One can see that among the clones we find a retrograde jovian co-orbital (green points) with $e$ and $i$ simultaneously near that of BZ509 (marked by the blue triangle in the figure). Figure~\ref{fig:clone_example} shows the evolution of this clone, which remains a jovian co-orbital for 3.5~Myr. It first reaches $i>90^o$ from the initial direct orbit after $\approx76$~kyr and then remains retrograde for the rest of its lifetime. The two 1:-1 resonant argument histories show periods of libration, of tens or hundreds of thousands of years duration, around $180^o$ or $0^o$, as well as there being $\omega$ oscillations around all of $0^o$, $90^o$, $180^o$ and $270^o$ at times. The red triangle at the bottom of each panel near 2.45~Myr indicates a time when $e$ and $i$ simultaneously match that of BZ509 (blue triangle in Fig.~\ref{fig:all_e_vs_i_clones}). \citet{Wiegertetal2017} demonstrate that BZ509 is currently librating in the same argument ($\phi^{\star}$ here) for at least the 20~kyr interval centered on the present day. Therefore, not only do we demonstrate a path from the main belt to a jovian co-orbital state with $e$ and $i$ simultaneous to BZ509, but also one that is librating around in $\phi^{\star}$ with the same $\approx 100^o$ libration amplitude.

In Figure~\ref{fig:all_e_vs_i_clones}, we see evidence of long-lived particles at high-$e$ and high-$i$ that can sit near a given $e$/$i$ for $\sim10$~Myr. Three examples are clear in Figure~\ref{fig:all_e_vs_i_clones}: one cluster of points is located at $e\approx0.65$ and $i\approx140^o$ (magenta), one at $e\approx0.95$ and $i\approx115^o$ (red), and one at $e\approx0.98$ and $i\approx105^o$ (yellow; see Appendix~\ref{sec:example long-lived} for the full orbital evolutions of these three long-lived retrograde objects with $a$ near $a_J$). Due to the long duration of these states, such long-lived objects are those most likely to be found; this is probably the context of the discovery of BZ509 on an orbit that stays near its current $e$ and $i$ (with a range of $e\approx0$, $i\approx145^o$ to $e\approx0.45$, $i\approx170^o$, paralleling the upper edge of the green dots in Figure~\ref{fig:all_e_vs_i_clones}) for at least $\approx200$~kyr \citep{Huangetal2018}. While it is possible but rare to reach the exact BZ509 state, other long-lived states near Jupiter exist (each single one of which is also rare); if the one known jovian long-lived retrograde co-orbital had been any of these, papers on its origin would have been written. We thus are unsure anything profound should be concluded from the particular current orbit of BZ509. (As an aside, it should be noticed that with a near-planar (albeit retrograde) and moderate eccentricity, BZ509 is the most detectable of the long-lived objects we illustrate.)

\section{Additional discussion}    \label{sec:discussion}

We thus believe there is a very plausible case that 2015 BZ$_{509}$ is an escaped main-belt asteroid that became retrograde in an already-demonstrated set of processes \citep{Greenstreetetal2012a, Greenstreetetal2012b, Granviketal2018} that happen in steady state. We have demonstrated a path from this source to an orbit nearly identical to BZ509; with the long-lived niches that last$>$1,000 times longer than the median trapping time, it is very likely that the object first found would be in such a niche state. Given the long duration of these states, most of the steady state retrograde `residence time' (that maps where the population is) shifts to these states, which we estimate are thus producing numbers that match in order of magnitude to the `statistics of one' example of BZ509. We here briefly discuss some other ideas for sources, and posit the idea that the main-belt path provided large-$i$ orbits to the outer Solar System as well.

The lack of {\it retrograde} jovian co-orbitals from our Centaur simulations might be due to the initial conditions of the incoming Centaurs in the simulations, which do not include the highest inclinations known to exist in the TNO scattering population (which feeds the Centaur population). It is not clear, however, whether the inclusion of these higher inclinations in the initial conditions would result in temporary capture into retrograde co-orbital motion at Jupiter since no estimate of the feeding efficiency from such a source to jovian retrograde orbits has ever been made. Our simulations do show the the (dominant) low-$i$ Centaur supply is not raised to high inclinations as they journey to lower $a$\footnote{\citet{HornerWynEvans2006} studied the capture of a sample of known Centaurs, which was biased toward the lowest-$a$, lowest-$i$ Centaurs by observation selection effects, into temporary co-orbital capture with the four giant planets. Their results revealed no retrograde co-orbital captures with any planet or the efficiency at which such transient co-orbital captures are made.}, and bringing $i>45^o$ TNOs to $a\simeq$~$a_J$ is very inefficient due to resultingly high planetary encounter speeds. Therefore, one would have to integrate an ensemble of large-$i$ TNOs to determine what (presumably small) fraction of them reach the long-lived jovian co-orbital state; this remains to be done but we suspect it will be many orders of magnitude rarer than the main-belt source.

\citet{NamouniMorais2018} numerically integrate one million clones of the nominal BZ509 orbit up to 4.5~Gyr into the past and suggest a 0.003\% chance that BZ509 would have $a$ near that of Jupiter 4.5~Gyr ago. Their Figure 3 (similar to our Fig.~\ref{fig:clone_example} above) shows $a$ stable near Jupiter 4.5~Gyr ago, but neither 1:-1 resonant argument shows libration in their figure. Of their remaining clones with $a\approx$~$a_J$, all but one have their $a$ increased above that of Jupiter's (and are not oscillating around Jupiter's semimajor axis). In fact, this `stable niche' is extremely similar to the case of Appendix Figure~\ref{fig:clone_example_2}, reached by our pre-retrograde cloning procedure from a main-belt source.

Fig.~\ref{fig:all_e_vs_i_clones} shows many retrograde objects with $a$ near Jupiter spending lots of time on near-polar ($i\approx90^o$) orbits. \citet{NamouniMorais2018} discuss what they call `the polar corridor' (with $i=90\pm45^o$ and semimajor axes of hundreds of thousands of AU); many of their backwards-integrated BZ509 clones spend tens of Myr `escaping' from the outer Solar System (where they then feel the galactic influence) via this corridor. They claim that a population of objects on near-polar orbits in the transneptunian object (TNO) and Centaur populations is evidence that these objects, including BZ509, originated from an extrasolar source since planet formation models of nearly-coplanar planetary orbits interacting with a coplanar planetesimal disk cannot produce large-inclination orbits stable on Gyr timescales. However, we show a clear path for asteroids coming out of the main belt in steady state reaching orbits with $i>90^o$ and $a$ near Jupiter's for timescales of order 10~Myr, removing this argument for a needing an extrasolar origin for these objects.

The generic issue with outer solar system origins (and the polar corridor specifically) is that {\it all} orbits eventually escape a meta-stable source and the giant planets eject essentially everything; when there is an {\it unbounded} phase space available the backwards integration then yields {\it no} estimate of the supply efficiency. To do this, one would have to integrate a huge set of inbound interstellar interlopers, having a range of impact parameters and drawn from the strongly hyperbolic inbound speed distribution, to determine what (presumably minuscule) fraction of them can reach the jovian co-orbital state; this remains to be done but we suspect it will be completely negligible compared to the main-belt source.

The polar corridor has another aspect, in the context of a few high-$i$ or retrograde transneptunian objects in the Minor Planet Center database with  $i>60^o$ and perihelion $q>15$~AU: 2002 XU$_{93}$, 2007 BP$_{102}$, 2008 KV$_{42}$ (nicknamed Drac; \citet{Gladmanetal2009}), 2010 WG$_9$, 2011 KT$_{19}$ (nicknamed Niku; \citet{Chenetal2016}), and 2014 LM$_{28}$. All of these objects have inclinations within $\pm30^o$ of a polar orbit at $i=90^o$; Drac ($i=103^o$) and Niku ($i=110^o$) are on retrograde ($i>90^o$) orbits. \citet{BatyginBrown2016a,BatyginBrown2016b} discuss the idea that these objects are populated by a hypothetical distant planet raising TNOs into the polar corridor, and perhaps from there they could reach a retrograde jovian co-orbital state. However, we again suspect the efficiency is extremely low because of the need to greatly lower the semimajor axes to that of Jupiter without the benefit of frequent and lower-speed encounters that the low-inclination state provides. The required demonstration is again simple in principle: If polar TNO orbits feed objects like BZ509, integrations from the estimated TNO region orbit distribution \citep{BatyginBrown2016b} can be forward propagated to estimate the steady-state number of jovian co-orbitals given the source population estimate. We note that this demonstration must not generate a very abundant polar Centaur population (with $5<a<30$~AU, $q>7.35$~AU) that violates survey constraints which have found very few of them (e.g., \citet{Petitetal2017}).

We actually here posit the {\it inverse} process: Could the population of objects in the outer Solar System on near-polar orbits have originated in the inner Solar System? After all, we have shown that the escaping NEA population already generates near-polar orbits and \citet{NamouniMorais2018} show these efficiently then populate the polar corridor and reach TNO semimajor axes. We thus have an existing Jupiter (not a hypothetical planet) that already creates and feeds large-$i$ orbits to the outer Solar System. The distance cut in our clone integrations at 19~AU prevents us from determining if we can produce particles with $a$ beyond Uranus, but we do find that 43\% of our clone particles are removed from the integrations for reaching heliocentric distances beyond the 19~AU cut. Could this be the origin of objects like Drac and Niku? Computed orbital evolutions in \citet{Gladmanetal2009} and \citet{Chenetal2016} show that these objects are metastable on Gyr timescales, so an {\it outflowing} (rather than incoming) polar corridor would result in Jupiter- and Saturn-crossing Centaurs to dominantly increase their semimajor axes until they reach orbits decoupled from the two most massive giants; once only Uranus and/or Neptune crossing, the dynamical lifetimes (and thus abundance in the steady state) become much larger. The depopulation of huge numbers of primordial inner-Solar system objects might be able to leave a surviving tail of high-$i$ TNOs beyound Jupiter, providing the postulated metastable source \citep{Gladmanetal2009} for the high-$i$ TNOs and Centaurs.

All of these options deserve quantitative exploration in future work. Given the information that we have, we favour the least dramatic hypothesis: that 2015 BZ$_{509}$ is a long-lived member of the known ensemble of high inclination orbits produced via leakage from the main asteroid belt.

\acknowledgments

S. Greenstreet acknowledges support from the Asteroid Institute, a program of B612, 20 Sunnyside Ave, Suite 427, Mill Valley, CA 94941. Major funding for the Asteroid Institute was generously provided by the W.K. Bowes Jr. Foundation and Steve https://iopscience.iop.org/article/10.1086/300720/fulltext/980143.text.htmlJurvetson. Research support is also provided from Founding and Asteroid Circle members K. Algeri-Wong, B. Anders, R. Armstrong, G. Baehr, The Barringer Crater Company, B. Burton, D. Carlson, S. Cerf, V. Cerf, Y. Chapman, J. Chervenak, D. Corrigan, E. Corrigan, A. Denton, E. Dyson, A. Eustace, S. Galitsky, L. \& A. Fritz, E. Gillum, L. Girand, Glaser Progress Foundation, D. Glasgow, A. Gleckler, J. Grimm, S. Grimm, G. Gruener, V. K. Hsu \& Sons Foundation Ltd., J. Huang, J. D. Jameson, J. Jameson, M. Jonsson Family Foundation, D. Kaiser, K. Kelley, S. Krausz, V. La\v{s}as, J. Leszczenski, D. Liddle, S. Mak, G.McAdoo, S. McGregor, J. Mercer, M. Mullenweg, D. Murphy, P. Norvig, S. Pishevar, R. Quindlen, N. Ramsey, P. Rawls Family Fund, R. Rothrock, E. Sahakian, R. Schweickart, A. Slater, Tito's Handmade Vodka, T. Trueman, F. B. Vaughn, R. C. Vaughn, B. Wheeler, Y. Wong, M. Wyndowe, and nine anonymous donors.

S. Greenstreet acknowledges the support from the University of Washington College of Arts and Sciences, Department of Astronomy, and the DIRAC Institute. The DIRAC Institute is supported through generous gifts from the Charles and Lisa Simonyi Fund for Arts and Sciences and the Washington Research Foundation. 

B. Gladman acknowledges support from the Natural Sciences and Engineering Research Council of Canada.

\appendix

\section{Example long-lived retrograde objects}
\label{sec:example long-lived}

\begin{figure}[h!]
\centering
\includegraphics[width=0.45\textwidth]{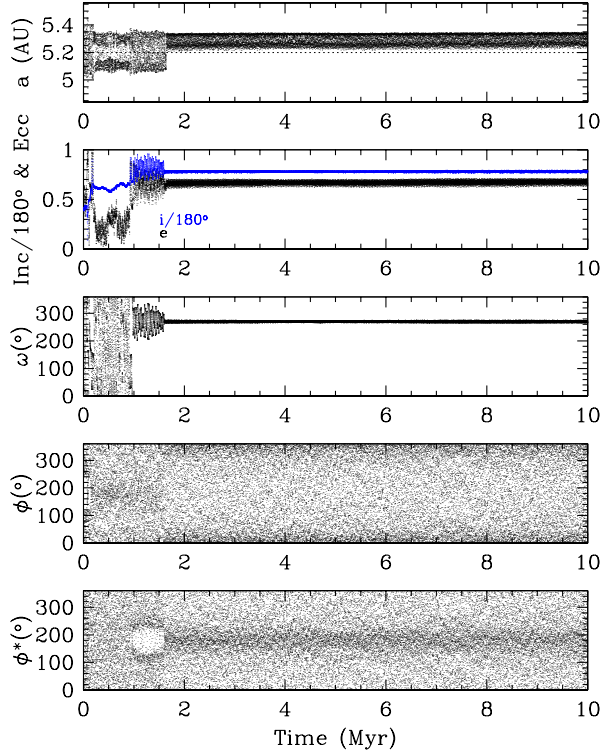}
\caption{Orbital history of the clone particle shown as the cluster of magenta dots at $e\approx0.65$ and $i\approx140^o$ in Figure~\ref{fig:all_e_vs_i_clones}. This particle is still alive at the end of the 10~Myr integration and spends its entire lifetime with $a$ near Jupiter's semimajor axis. The particle flips to a retrograde orbit at $\approx20$~kyr into its lifetime. Coupled Kozai $e$ and $i$ oscillations start around $900$~kyr into the integration with $e\approx0.68$ and $i\approx140^o$ while the argument of pericenter $\omega$ librates around $270^o$. $\approx700$~kyr later, these librations tighten. The resonant arguments $\phi_{1:-1}=\lambda-\lambda_J$ and $\phi^\star_{1:-1}=\lambda-\lambda_J-2\omega$ show resonant behavior (libration around $0^o$) both before and after the retrograde flip, indicating the particle is at times formally in the 1:-1 co-orbital resonance with Jupiter. 
\label{fig:clone_example_2}}
\end{figure} 

\begin{figure}[h!]
\centering
\includegraphics[width=0.45\textwidth]{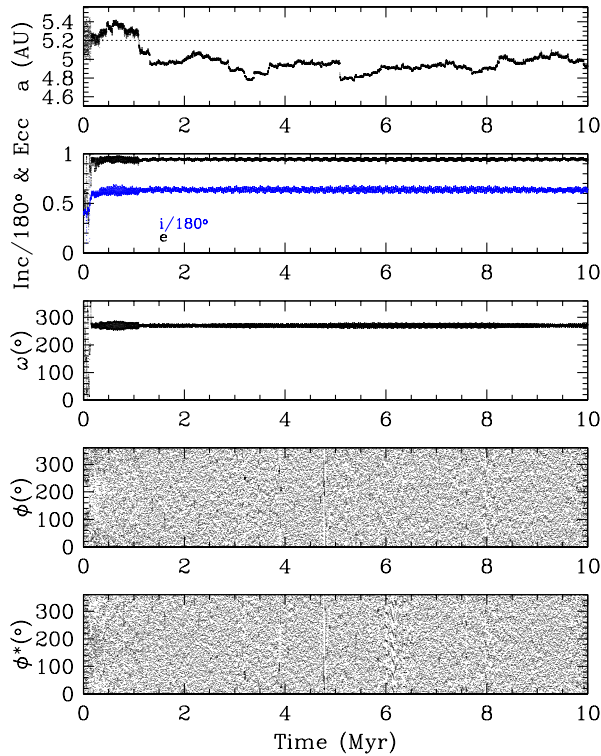}
\caption{Orbital history of the clone particle with $e$ and $i$ concentrated around $e\approx0.95$ and $i\approx115^o$ (shown in red) in Figure~\ref{fig:all_e_vs_i_clones}. This particle lives for the full 10~Myr integration and spends its lifetime at $a$ within 0.4~AU of Jupiter's semimajor axis, mostly interior to $a_J$. The particle flips to a retrograde orbit at $\approx250$~kyr into its lifetime. Coupled Kozai oscillations in $e$ and $i$ start almost immediately following the flip with $e\approx0.95$ and $i\approx115^o$ (which accounts for the concentration of red dots in Figure~\ref{fig:all_e_vs_i_clones} at this $e$ and $i$) while the argument of pericenter $\omega$ librates around $270^o$. These librations remain tightly bound to this $e$, $i$, and $\omega$ for 10~Myr. The resonant argument $\phi_{1:-1}=\lambda-\lambda_J$ shows resonant behavior (libration around $0^o$) for only a brief $\approx100$~kyr period at the start of the integration. 
\label{fig:clone_example_3}}
\end{figure}

\begin{figure}[h!]
\centering
\includegraphics[width=0.45\textwidth]{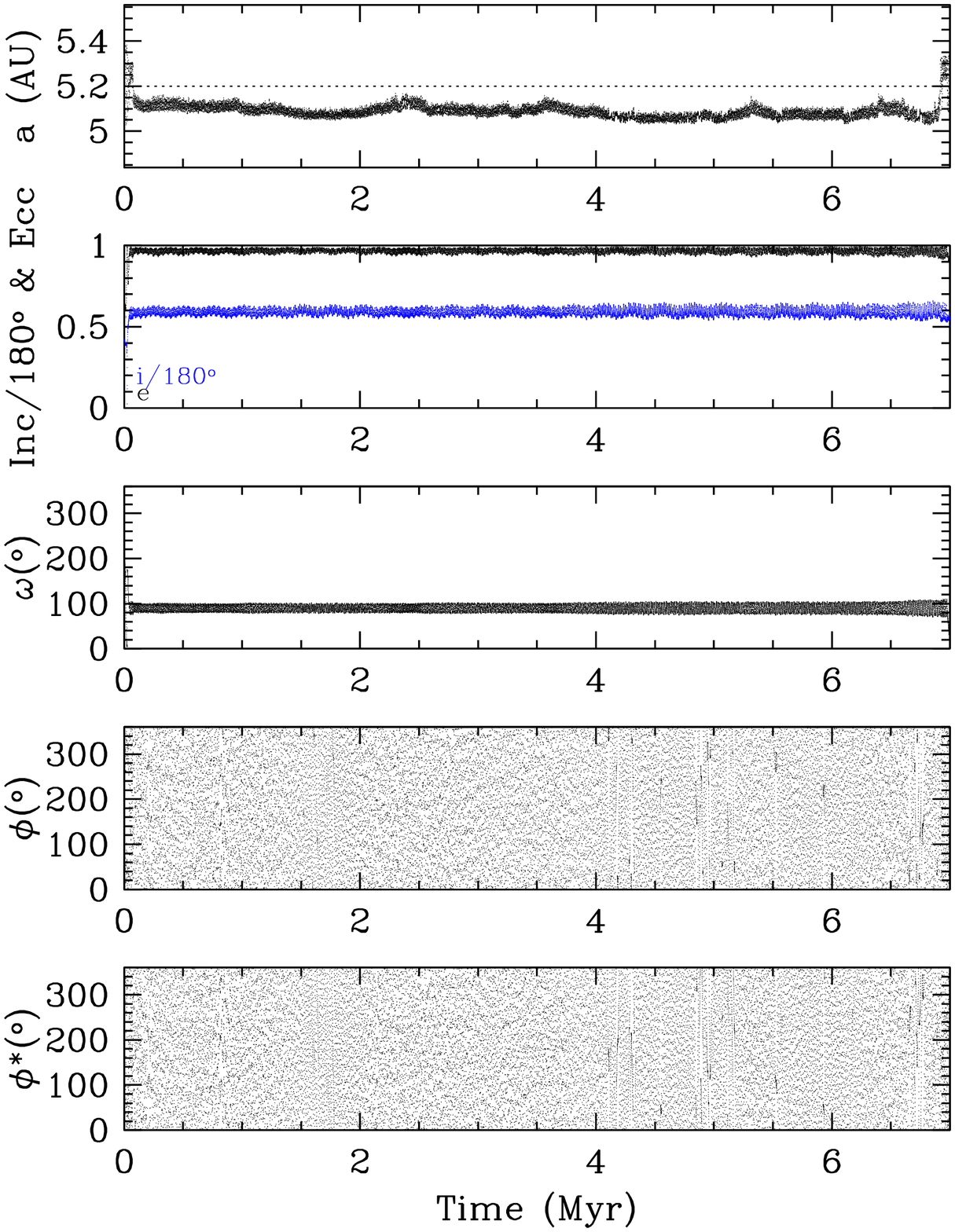}
\caption{Orbital history of the clone particle depicted in Figure~\ref{fig:all_e_vs_i_clones} as the grouping of yellow dots at $e\approx0.98$ and $i\approx110^o$. This particle is removed from the integration at $\approx7$~Myr when $e$ goes to 1 and the particle is pushed into the Sun. This particle spends almost the entirety (6.9~Myr) of its lifetime with $a$ just interior to and within 0.2~AU of $a_J$; the particle's $a$ only oscillates around $a_J$ for the first $\approx80$~kyr and the last $\approx20$~kyr of its lifetime. The particle's inclination evolves to $i>90^o$ at $\approx30$~kyr into its lifetime. Coupled Kozai $e$ and $i$ oscillations start immediately thereafter and remain locked with $e\approx0.98$ and $i\approx110^o$ while the argument of pericenter $\omega$ likewise librates around $90^o$ for the remainder of the particle's lifetime. The resonant arguments $\phi_{1:-1}=\lambda-\lambda_J$ and $\phi^\star_{1:-1}=\lambda-\lambda_J-2\omega$ both show libration around $0^o$ for tens of kyr at the very beginning and end of the particle's lifetime (corresponding to the interval where the semimajor axis rapidly changes).
\label{fig:clone_example_4}}
\end{figure}
 
We selected one of the eight initial particles in our NEO model that became retrograde and cloned it just after it reached the near-Jupiter state, but before it became retrograde (Figure~\ref{fig:a_e_i_NEO_retro_and_zoom}). The goal was to determine if reaching the retrograde state was then common and whether any particles that did so exhibited long-lived states and/or passed near the ($e,i$) state of BZ509. Figures~\ref{fig:clone_example_2}, \ref{fig:clone_example_3}, and \ref{fig:clone_example_4} show orbital evolutions of three such long-lived retrograde clones, that show up as clusters of points in magenta, red, and yellow, respectively, in Figure~\ref{fig:all_e_vs_i_clones}. Two of these particles (those in Figures~\ref{fig:clone_example_2} and \ref{fig:clone_example_3}) are still alive at the end of the 10~Myr integration. The particle shown in Figure~\ref{fig:clone_example_4} is removed from the integration after $\approx7$~Myr when it is pushed into the Sun.

Each of these particles depart from the state where their semimajor axes are oscillating around that of Jupiter early in their clone lifetimes. The particle shown in Figure~\ref{fig:clone_example_2} has its $a$ evolve to just outside $a_J$, oscillating between $a\approx5.2-5.35$~AU for the last $\approx8.2$~Myr of its lifetime. The particles shown in Figures~\ref{fig:clone_example_3} and \ref{fig:clone_example_4} have their semimajor axes evolve to interior to $a_J$. Figure~\ref{fig:clone_example_3} shows $a$ dropping as low as 4.8~AU, while the particle in Figure~\ref{fig:clone_example_4} remains within 0.2~AU of $a_J$ for the remainder of its $\approx7$~Myr lifetime. 

The inclinations for these three particles all become retrograde quickly, where they remain. The eccentricity, inclination, and argument of pericenter for these particles remain surprisingly constant for the remainder of their integrations; we have confirmed that this is classic Kozai behavior (the $e$ and $i$ oscillations both being coupled to the phase of the small $\omega$ libration). The resulting small $e,i$ variations create the clusters of points seen in Figure~\ref{fig:all_e_vs_i_clones}. The two 1:-1 resonant argument histories are shown in the bottom two panels of each figure. $\phi$ shows periods of libration around $0^o$ at the start of evolution in all three cases. $\phi^{\star}$ shows libration around $0^o$ from $\approx1-1.6$~Myr for the particle in Figure~\ref{fig:clone_example_2} and for the first $\approx40$~kyr of the lifetime of the particle in Figure~\ref{fig:clone_example_4}. It almost seems that the 1:-1 librations are responsible for feeding the particles from a fully resonant state to this Kozai lock but out of formal resonance. Once the latter is established, it is the Kozai lock that stabilizes the orbit, for with $a\simeq a_J$ the large eccentricity means that both nodes are considerably closer to the Sun and no jovian close encounters can occur. Were these objects to ever come close to the terrestrial planets, the large orbital inclination results in such high encounter speeds that only small semimajor axis changes could occur.




\clearpage

\end{document}